\documentclass[12pt]{article}
\usepackage{latexsym}
\usepackage{graphicx}
\usepackage[section]{placeins}

\newbox\SlashedBox  
\def\fs#1{\setbox\SlashedBox=\hbox{#1} 
\hbox to 0pt{\hbox to 1\wd\SlashedBox{\hfil/\hfil}\hss}{#1}} 
\def\hboxtosizeof#1#2{\setbox\SlashedBox=\hbox{#1} 
\hbox to 1\wd\SlashedBox{#2}} 

\def\ms#1{\setbox\SlashedBox=\hbox{$#1$}
\hbox to 0pt{\hbox to 1\wd\SlashedBox{\hfil/\hfil}\hss}#1}

\newcommand{\tr}{{\rm tr}}
\newcommand{\ie}{{\em i.e.~}}

\newcommand{\be}{\begin{equation}}
\newcommand{\ee}{\end{equation}}
\newcommand{\ba}{\begin{eqnarray}}
\newcommand{\ea}{\end{eqnarray}}

\begin{document} 

\thispagestyle{empty}

\begin{flushright}
ROM2F/00/6 \\
DAMTP-2000-34
\end{flushright}

\vspace{1.5cm}

\begin{center}

{\LARGE {\bf Anomalous dimensions  \\ 
in ${\cal N}$=4 SYM theory at order $g^{4}$ \rule{0pt}{25pt} }} \\
\vspace{1cm} \
{Massimo Bianchi$^{\:a}$, Stefano Kovacs$^{\:b}$, 
Giancarlo Rossi$^{\:a}$ and Yassen  S. Stanev$^{\:a\,\dagger}$} \\ 
\vspace{0.6cm} 
{$^a$ {\it Dipartimento di Fisica, \ Universit{\`a} di Roma \  
``Tor Vergata''}} \\  {{\it I.N.F.N.\ -\ Sezione di Roma \ 
``Tor Vergata''}} \\ {{\it Via della Ricerca  Scientifica, 1}} 
\\ {{\it 00173 \ Roma, \ ITALY}} \\ 
\vspace*{0.5cm}
{$^b$ {\it D.A.M.T.P., University of Cambridge}} \\
{{\it Wilberforce Road, Cambridge, CB3 0WA, UK}}

\end{center}

\vspace{1cm}

\begin{abstract}

We compute four-point correlation functions
of scalar composite operators in the ${\cal N}$=4 supercurrent 
multiplet at order $g^{4}$ using the ${\cal N}$=1 superfield formalism.
We confirm the interpretation of short-distance logarithmic behaviours
in terms of anomalous dimensions of unprotected operators exchanged 
in  the intermediate channels and we determine the two-loop 
contribution to the anomalous dimension of the ${\cal N}$=4 Konishi
supermultiplet.

\end{abstract}

\vspace{4cm}
\noindent
\rule{6.5cm}{0.4pt} 

{\footnotesize ${}^{\dagger}$~On leave of absence from Institute for 
Nuclear Research and Nuclear Energy, Bulgarian Academy of Sciences, 
BG-1784, Sofia, Bulgaria}
\newpage 

\setcounter{page}{1}

\section{Introduction and summary of the results}

The conjectured AdS/CFT correspondence~\cite{jm,gkp,wads,agmoo} has
renewed 
the interest in conformal field theories (CFT's), in particular 
${\cal N}$=4 supersymmetric Yang-Mills theory (SYM) in $D=4$, and has 
raised the issue
of finding to what extent high order 
perturbative computations are feasible in the weak coupling regime.

The interest is twofold. On the one hand, given the explicit analytic 
expressions of certain four-point amplitudes in the AdS context, 
one may ask whether it is possible to recognize some systematic 
pattern 
in the perturbative ${\cal N}$=4 field-theoretic results. 
On the other hand, insights gained in such finite 
theories may prove to be useful in theories with (partial) 
supersymmetry breaking. 

In CFT the dependence of two- and three-point functions of scalar
primary operators on the external insertion points is completely 
fixed by conformal symmetry. For protected operators 
non-renormalisation theorems hold that prevent quantum corrections 
to the lowest order contribution. Explicit computations have 
confirmed the validity of these theorems in the context of the AdS/CFT
correspondence \cite{lmrs,dfs,skiba}.
Except for extremal \cite{extremal,bkextr,ehsswextr} and sub-extremal 
correlators \cite{epv}, no such non-renormalisation theorems are
believed to hold for higher-point functions.

A particularly interesting class of higher-point functions are the
four-point functions of the lowest chiral primary operators, \ie 
dimension-two gauge-invariant scalar composites in the ${\cal N}$=4 
supercurrent multiplet. 
Perturbative computations involving such operators have 
been performed at order $g^{2}$ in~\cite{ehssw,bkrs}.
Motivated by the presence of string corrections to the AdS effective 
action~\cite{gg,bg}, instanton computations have been carried out  
in~\cite{bgkr,como,dhkmv,bkrs}.
All such computations show logarithmic behaviours at short-distances 
that allow an interpretation in terms of anomalous dimensions of 
unprotected operators in long supermultiplets exchanged in 
intermediate channels~\cite{bkrs}. Similar analyses have been 
performed in the AdS context in the supergravity 
limit~\cite{dmmrope}. 
A detailed comparison of the two classes of results is problematic 
because of the different regimes 
in the ($g^{2}$, $N$) parameter space that are explored by the two 
approaches. Nevertheless, explicit AdS
computations in the scalar sector that are amenable to a
perturbative analysis in the field-theoretic weak coupling regime, 
such as the one presented here, have been recently completed in~\cite{arfr2} 
in the supergravity approximation with the help of the results 
of~\cite{arfr1}.

In this paper, we compute  at order $g^{4}$ four-point correlation functions
of the lowest dimension scalar composite operators in the ${\cal N}$=4 
supercurrent multiplet using the ${\cal N}$=1 superfield formalism.
We confirm the interpretation of the short-distance logarithmic behaviour
in terms of anomalous dimensions of unprotected operators exchanged in 
intermediate channels and we extract the two-loop contribution to the 
anomalous dimension of the ${\cal N}$=4 Konishi supermultiplet. 
In order to illustrate our point we will concentrate in most of our 
presentation on the simplest of the six independent four-point 
functions of 
chiral primary operators~\cite{ehssw,bkrs} (see below). Identities 
that have been proved to hold both in perturbation theory 
in~\cite{ehssw,ehpsw} and at the 
one-instanton level in~\cite{bkrs} should be enough to determine
all four-point functions from the one we consider. 

A recent interesting paper by Eden, 
Schubert and Sokatchev has reported on similar computations at order
$g^{4}$ from 
a different vantage point and confirmed the validity of the above 
identities~\cite{ess}.  
Their approach is based on the less familiar ${\cal N}$=2 harmonic 
superspace formalism. It is remarkable that, although we adopt a 
different approach, we find the same result for the four-point 
function on which we focus our attention, up to an overall constant 
that was not fixed in~\cite{ess}.
We would like to stress that, contrary to expectations, 
the ${\cal N}$=1 superfield approach we have pursued is not 
prohibitively complicated and it is of wider applicability.
In particular we have in mind applications to 
some interesting finite ${\cal N}$=1
superconformal theories that are promising candidates for realistic 
theories after soft supersymmetry breaking.

The plan of the paper is as follows. In section 2, 
for the sake of completeness, we will write down the action, 
propagators and vertices in the ${\cal N}$=1 superfield formalism and 
identify the six independent four-point functions of chiral primary 
operators. In section 3 we draw the relevant 
Feynman 
superdiagrams and sketch the computation for the simplest possible 
four-point function. In section 4 we discuss the final result and 
interpret the dominant logarithmic behaviours at short distance in 
terms of the anomalous dimension of the lowest dimensional operator, 
 ${\cal K}_{\bf 1}$, in 
the ${\cal N}$=4 Konishi supermultiplet. 
We confirm the one-loop results of~\cite{bkrs} and
extract the two-loop contribution\footnote{Notice that we call 
 ``$\ell$-loops'' calculations that are of order ${g^{2\ell}}$. The 
authors of refs.~\cite{ehssw} and~\cite{ess} dub diagrams of this 
order as $\ell+1$-loop calculations.} to the anomalous dimension of 
${\cal K}_{\bf 1}$. We finally comment on possible extensions 
of our results to finite ${\cal N}$=1 supersymmetric theories that 
arise after soft breaking of ${\cal N}$=4 supersymmetry.

\section{${\cal N}$=4 SYM in ${\cal N}$=1 superspace}

The field content of ${\cal N}$=4 SYM \cite{n4sym} comprises a 
vector, $A_{\mu}$, four Weyl spinors, $\psi^{A}$ ($A$=1,2,3,4), and 
six real scalars, $\varphi^{i}$ ($i$=1,2,\ldots,6),
all in the adjoint representation of the gauge group ${\cal G}$. 
Since no off-shell formulation is available that manifestly preserves 
${\cal N}$=4 supersymmetry, in order to compute quantum corrections  
one has to resort either to the ${\cal N}$=2 harmonic superspace approach 
pursued in~\cite{ehssw,ess} or to the more familiar ${\cal N}$=1 
formalism pursued in~\cite{bkrs}. 

In the ${\cal N}$=1 formalism the fundamental fields can be arranged
into a vector superfield, $V$, and three chiral superfields, 
$\Phi^{I}$ ($I$=1,2,3). The six real scalars, $\varphi^{i}$, are 
combined into three complex fields, namely 
\begin{equation}
    \phi^{I} = \frac{1}{\sqrt{2}} \left( \varphi^{I}+i\varphi^{I+3} 
    \right) \:  \qquad \phi^{\dagger}_{I} = \frac{1}{\sqrt{2}} \left( 
    \varphi^{I}-i\varphi^{I+3} \right) \; ,
    \label{defphi}
\end{equation}
that are the lowest components of the superfields $\Phi^{I}$ and 
$\Phi^{\dagger}_{I}$, respectively. Three 
of the Weyl fermions, $\psi^{I}$, are the spinors of the 
chiral multiplets. The fourth spinor, $\lambda = \psi^{4}$, together 
with the vector, $A_{\mu}$, form the vector multiplet. 
In this formulation only a $SU(3)\times U(1)$ subgroup of the 
original $SU(4)$ R-symmetry group is manifest. $\Phi^{I}$ and 
$\Phi^{\dagger}_{I}$ transform in the representations {\bf 3} and 
${\overline {\bf 3}}$ of the global $SU(3)$ ``flavour'' Q-symmetry, 
while $V$ is a singlet. 
Under the axial $U(1)$ R-symmetry the vector $A_{\mu}$ is neutral,
the gaugino  $\lambda$ has charge $+3/2$,
the spinors of the chiral multiplets $\psi^{I}$ have charge $-1/2$ 
and the three complex scalars $\phi^{I}$ have charge $+1$.

The (Euclidean) action in the ${\cal N}$=1 superfield formulation 
reads
\begin{eqnarray}
    S & = & {-2}\, \tr \left\{ \int d^{4}x \, 
    \left[ 
    \left( 
    \int d^{2}\theta \: \frac{1}{16} W^{\alpha}W_{\alpha} + {\rm h.c.}
    \right) 
    + \left( \int d^{2}\theta d^{2}\bar\theta \, 
    e^{-{gV}} \Phi^{\dagger}_{I}
    e^{{gV}}\Phi^{I} \right) \right. \right. \nonumber \\
    & - & \left. \left. \!\!\frac{g}{3!\sqrt{2}} \left( \int 
d^{2}\theta \,  
    \varepsilon_{IJK} \Phi^{I}[\Phi^{J},\Phi^{K}] - \int 
d^{2}{\overline 
    \theta} \, \varepsilon^{IJK}
    \Phi^{\dagger}_{I}[\Phi^{\dagger}_{J},\Phi^{\dagger}_{K}] 
    \rule{0pt}{18pt} \right) \right] \right\} \; ,
    \label{n1superfield}
\end{eqnarray}
where $W_{\alpha}$ is the chiral superfield-strength of 
$V$ 
\be
W_{\alpha}= -{1 \over 4 g} \bar {D^{2}} \left( e^{-{2gV}} D_{\alpha} 
 e^{{2gV}} \right) \, .
\ee
The trace over the colour indices is defined by 
\be
\tr(T^{a} T^{b}) = {1 \over 2 } \delta^{ab} \quad a,b = 
1,\ldots, {\mbox{dim}}({\cal G}) \, ,
\label{tracedef}
\ee
where $T^{a}$ are the generators in the fundamental representation of 
the gauge group.  

Notice that a gauge fixing term, $S_{{\rm gf}}$, has to be added to
the classical action~(\ref{n1superfield}). As usual we 
decide to take 
\begin{equation}
    S_{{\rm gf}} =  \frac{1}{16 \alpha} \tr \int d^{4}x \int 
    d^{2}\theta\,d^{2}{\overline\theta} \left[ 
    \left({ D}^{2}V\right)\left({\bar { D}}^{2}V\right) 
    \right] \; ,
    \label{gaugefix}
\end{equation}
where $\alpha$ is a gauge parameter. We will not display here 
ghost terms since they do not contribute to the Green 
functions that we will consider at the order we work. 
The choice $\alpha = 1$ greatly simplifies all the 
computations, as it makes all corrections to the 
propagators of the fundamental superfields vanishing at order 
$g^{2}$~\cite{finite,kov}.  

Expanding the exponentials $e^{\pm gV}$ in~(\ref{n1superfield}) gives 
\begin{eqnarray}
    S &=& \int d^{4}x\:d^{2}\theta d^{2}\bar\theta\, 
\left\{\rule{0pt}{18pt} 
    V^{a} \Box  V_{a} 
    - \Phi^{a\dagger}_{I}\Phi^{I}_{a} - ig 
f_{abc}{\Phi^{\dagger}}^{a}_{I}
    V^{b}\Phi^{Ic} + {g^{2}\over 2} f_{abe} 
f_{ecd}{\Phi^{\dagger}}^{a}_{I}
    V^{b} V^{c} \Phi^{Id}  \right.  \nonumber \\ 
    && \left. - \frac{ig}{3!\sqrt{2}} f^{abc} \left[ 
\varepsilon_{IJK}
    \Phi_{a}^{I} \Phi_{b}^{J} \Phi_{c}^{K} \delta({\overline \theta}) 
- 
    \varepsilon^{IJK} \Phi^{\dagger}_{aI}\Phi^{\dagger}_{bJ}
    \Phi^{\dagger}_{cK}\delta(\theta) \right] + 
    \ldots \right\} \; ,
\label{actionsuper}	
\end{eqnarray}
where $f_{abc}$ are the structure constants of the gauge group 
and we have displayed 
only the terms that are relevant for our subsequent computations.

In the following we will carry on the calculations using ${\cal
N}$=1 formalism in coordinate superspace, that is more suitable for the study of
CFT's. The superfield propagators in the conventions 
of eq.~(\ref{actionsuper}) are 
\be
\langle \Phi^{\dagger}_{Ia} (x_i,\theta_i, \bar \theta_i)
\Phi^J_b (x_j,\theta_j, \bar \theta_j) \rangle =
- {{\delta_{I}}^{J} \delta_{ab} \over 4 \pi^2}
e^{i\left( \xi_{ii} +\xi_{jj} -2 \xi_{ji} \right) \cdot \partial_i}  
{1 
\over x_{ij}^2}
\label{propfi}
\ee
 
\be
\langle V_a (x_i,\theta_i, \bar \theta_i)
V_b (x_j,\theta_j, \bar \theta_j) \rangle =
{\delta_{ab} \over 8 \pi^2}
{\delta(\theta_{ij}) \delta(\bar \theta_{ij}) \over x_{ij}^2} \; ,
\label{propv}
\ee
where
\be
x_{ij} = x_{i}-x_{j} \qquad \theta_{ij}= \theta_{i}- \theta_{j}
\qquad \xi^{\mu}_{ij}= \theta_i^{\alpha} \sigma^{\mu}_{\alpha {\dot 
\alpha} } \bar \theta_j^{\dot \alpha} \: .
\ee

In the next section we will describe the calculation of the order $g^{4}$ 
perturbative correction to the four-point correlation function
\begin{equation}
	G^{(H)}(x_{1}, x_{2}, x_{3}, x_{4}) = 
	\langle {\cal C}^{11}(x_1)  {\cal C}^\dagger_{11}(x_2)  
	{\cal C}^{22}(x_3) {\cal C}^\dagger_{22}(x_4)\rangle \; 
	\label{h4point} \: ,
\end{equation}
where the gauge-invariant composite operators
\begin{equation}
{\cal C}^{IJ} = \tr (\phi^{I}\phi^{J}) \quad \quad  
{\cal C}^\dagger_{IJ} = \tr (\phi^{\dagger}_{I}\phi^{\dagger}_{J}) 
\label{defchir}
\end{equation}
are the lowest components of the (anti-)chiral superfields 
\begin{equation}
{C}^{IJ} = \tr (\Phi^{I}\Phi^{J}) \quad \quad  
{C}^\dagger_{IJ} = \tr (\Phi^{\dagger}_{I}\Phi^{\dagger}_{J}) 
\; .
\label{defsuperchir}
\end{equation}
In turn ${\cal C}^{IJ}$ and ${\cal C}^\dagger_{IJ}$ appear in the 
decomposition of ${\cal 
Q}_{{\bf 20}}^{ij}$, the lowest scalar components of the ${\cal N}$=4 
current supermultiplet, under $SU(4)\rightarrow SU(3)\times U(1)$. 
The ${\cal Q}_{{\bf 20}}^{ij}$ belong to the real representation ${\bf 
20}$ of $SU(4)$ and are defined as  
\begin{equation}
{\cal Q}_{{\bf 20}}^{ij} = \tr( \varphi^i \varphi^j - {\delta^{ij} 
\over 6} 
\varphi_k \varphi^k) \; . 
\end{equation} 

The most general four-point function of the ${\cal Q}_{{\bf 20}}^{ij}$
\begin{equation}
	G^{({\cal Q})}(x_{1}, x_{2}, x_{3}, x_{4}) = 
	\langle {\cal Q}^{i_{1}j_{1}}(x_1)  {\cal Q}^{i_{2}j_{2}}(x_2)  
	{\cal Q}^{i_{3}j_{3}}(x_3) {\cal Q}^{i_{4}j_{4}}(x_4)\rangle \; 
	\label{q4point} 
\end{equation}
can be expressed as a linear combination of $G^{(H)}$, defined in 
 eq.~(\ref{h4point}), and the following five independent four-point 
functions 
\ba
G^{(V)}(x_{1}, x_{2}, x_{3}, x_{4}) &=& 
	\langle {\cal C}^{11}(x_1)  {\cal C}^\dagger_{11}(x_2)  
	{\cal C}^{11}(x_3) {\cal C}^\dagger_{11}(x_4)\rangle \; 
	\label{all4point}
	\\
G^{(1)}(x_{1}, x_{2}, x_{3}, x_{4}) &=&
	\langle {\cal C}^{11}(x_1)  {\cal C}^\dagger_{22}(x_2)  
	{\cal C}^{22}(x_3) {\cal C}^\dagger_{11}(x_4)\rangle \; \nonumber \\ 
G^{(2)}(x_{1}, x_{2}, x_{3}, x_{4}) &=&
	\langle {\cal C}^{11}(x_1)  {\cal C}^{22}(x_2)  
	{\cal C}^\dagger_{11}(x_3) {\cal C}^\dagger_{22}(x_4)\rangle \; 
	\nonumber \\
G^{(3)}(x_{1}, x_{2}, x_{3}, x_{4}) &=&
	\langle {\cal C}^{11}(x_1)  {\cal C}^\dagger_{11}(x_2)  
	{\cal C}^{11}(x_3) {\cal C}^\dagger_{11}(x_4)\rangle \; \nonumber \\ 
G^{(4)}(x_{1}, x_{2}, x_{3}, x_{4}) &=&
	\langle {\cal C}^{11}(x_1)  {\cal C}^{11}(x_2)  
	{\cal C}^\dagger_{11}(x_3) {\cal C}^\dagger_{11}(x_4)\rangle 
	\nonumber \; .
\ea

We stress that the above five linearly independent four-point 
functions have been shown to be functionally related 
to~(\ref{h4point}) both in perturbation theory~\cite{ehssw,ehpsw,ess} 
and non-perturbatively at the one-instanton 
level~\cite{bkrs}.
For instance one finds 
\be 
G^{(H)}(x_{1}, x_{2}, x_{3}, x_{4}) = 
s \, G^{(V)}(x_{1}, x_{2}, x_{3}, x_{4}) \label{gvgh}
\ee
where $s$ is one of the two 
independent conformally invariant cross-ratios 
\be
	r = {x_{12}^{2}x_{34}^{2} \over x_{13}^{2}x_{24}^{2}}
	\;  , \quad
	s = {x_{14}^{2}x_{23}^{2} \over x_{13}^{2}x_{24}^{2}}
\label{crossrat}
\ee
that can be constructed out of four points.

\section{Superdiagrams and calculations}

The four-point function $G^{(H)}(x_1,x_2,x_3,x_4)$ defined 
in~(\ref{h4point}) is the lowest component of the supercorrelation 
function
\be
\Gamma^{(H)}(z_1,z_2,z_3,z_4) = \langle {C}^{11}(z_1)  
{C}^{\dagger}_{11}(z_2)  
C^{22}(z_3)  
C^{\dagger}_{22}(z_4) \rangle
\label{4pointsuper}
\ee
{\it viz.}
\be
G^{(H)}(x_1,x_2,x_3,x_4) = \Gamma^{(H)}(z_1,z_2,z_3,z_4) 
\vert_{\theta_{i}={\overline\theta}_i=0} \: ,
\ee
where $z_{i}=(x^{\mu}_{i}, \theta^{\alpha}_{i}, 
\bar\theta_{i\dot\alpha})$ as usual.
The ${\cal N}$=1 superfield approach greatly simplifies the 
calculations. The choice of
external flavours guarantees that there are no connected supergraphs 
at tree level. The potential corrections to the propagators at 
order $g^{4}$ would only contribute to disconnected supergraphs~\footnote{
Order $g^{4}$ corrections to 
two-point functions of operators belonging to ultra-short
supermultiplets vanish~\cite{psz}.} since the choice $\alpha =1$ in 
the gauge-fixing term makes all 
propagator corrections vanish at order $g^{2}$~\cite{finite,kov}.

\begin{figure}[!htbp]
 \begin{minipage}[t]{\linewidth}
	\centering
	\includegraphics[width=0.85\linewidth]{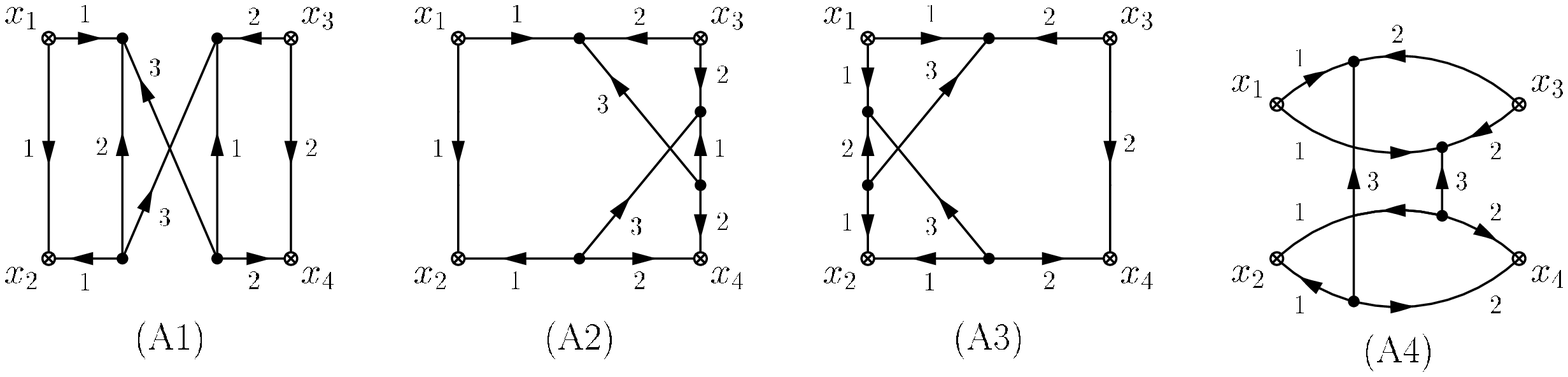}
	\caption{Diagrams with only chiral lines.}
	\label{diagrA}
 \end{minipage} \\ [2.5cm]
 \begin{minipage}[b]{\linewidth}
	\centering
	\includegraphics[width=0.85\linewidth]{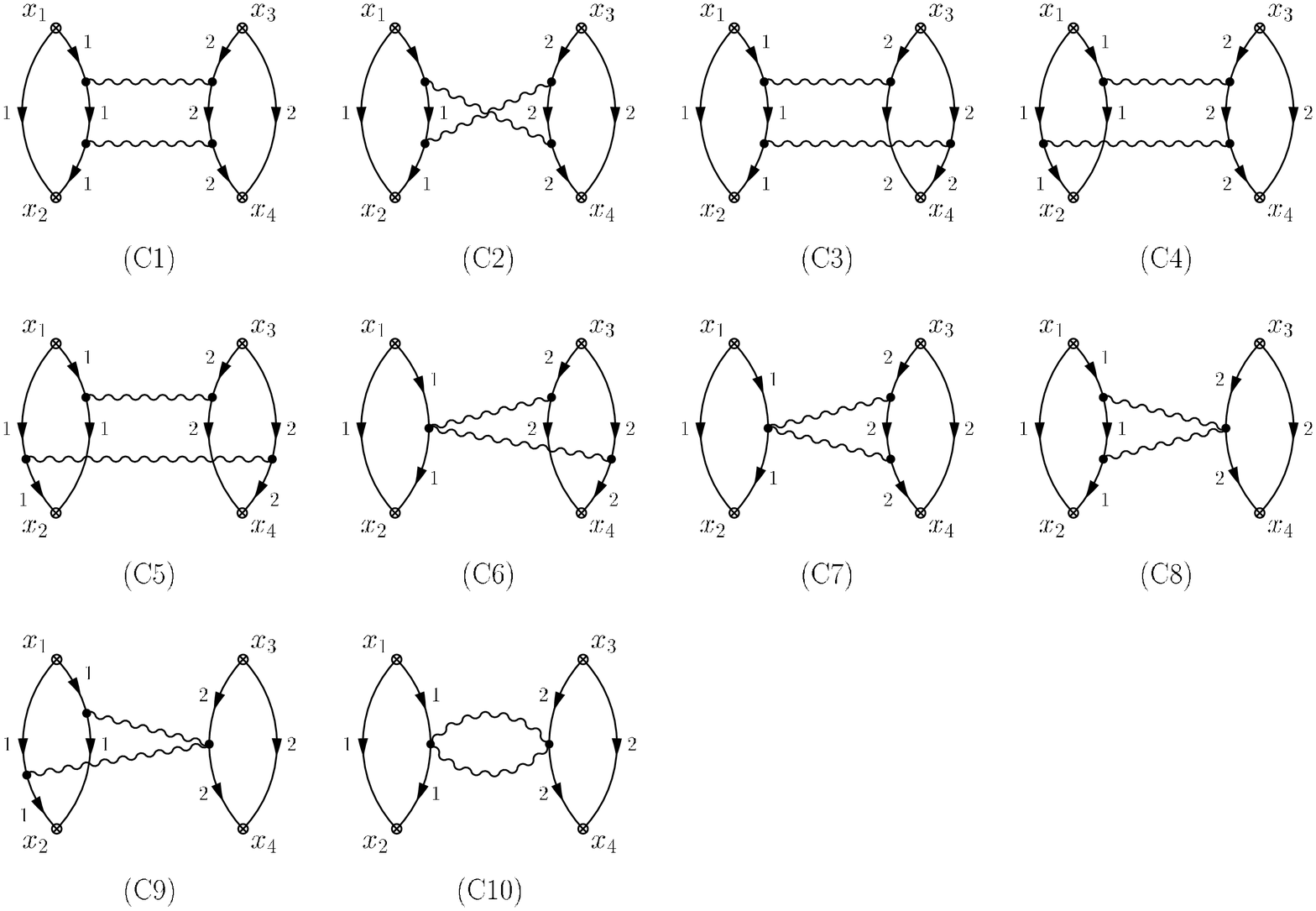}
	\caption{Diagrams with two vector lines.}
	\label{diagrC}
 \end{minipage}
\end{figure}

\begin{figure}[!htbp]
	\centering
	\includegraphics[width=0.85\linewidth]{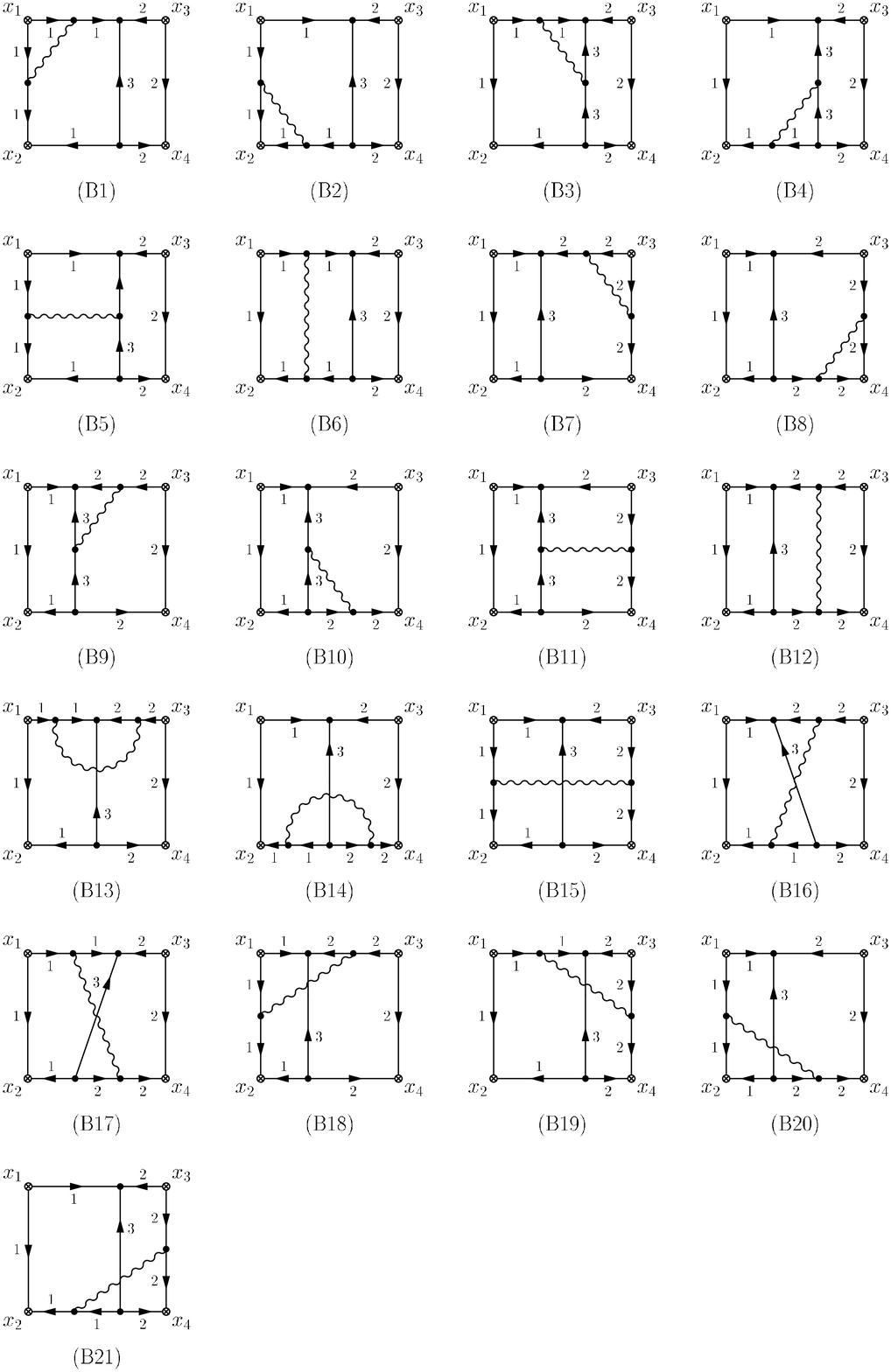}
	\caption{Diagrams with one vector line.}
	\label{diagrB}
\end{figure}

The connected superdiagrams contributing to~(\ref{4pointsuper}) 
are reported in Figs.~\ref{diagrA}, \ref{diagrC} and~\ref{diagrB}. 
We have only displayed superdiagrams that do not vanish by colour 
contractions. There are 4 superdiagrams
with only chiral lines, 21 superdiagrams with one internal vector line
and 10  superdiagrams with two internal vector lines. Notice that none
of them involves cubic or quartic pure vector vertices. This 
explains why we have not explicitly displayed the corresponding 
couplings in~(\ref{actionsuper}).

The computation of the overall 
weights, due to combinatorial factors and colour and flavour 
contractions, is greatly simplified if fake different 
coupling constants are introduced for each interaction term in the action. 
Only at the very end one imposes the relations among the couplings
that guarantee ${\cal N}$=4 supersymmetry.
As a result one finds\footnote{In what follows we will only display 
group theory factors relevant for the case ${\cal G}= SU(N)$. The  
generalization to an arbitrary gauge group amounts to substituting $g^{2}N$ with 
$g^{2}C_{2}(A)$ and $N^{2}-1$ with ${\mbox{dim}}({\cal G})$, 
$C_{2}(A)$ being the quadratic Casimir of the adjoint representation. } 
for the superdiagrams with only chiral lines
\be
A_{\rm tot} = {g^4 \over 8} N^2(N^2-1)  \left( 2 A_{1} + A_{2} + A_{3} + 
A_{4}\right)  \; ,
\label{0vweights}
\ee
for those with one vector line 
\ba
B_{\rm tot} &=& -{g^4 \over 4} N^2(N^2-1) \left(   
2 B_{1} +2 B_{2} +B_{3}+B_{4}-B_{5}-2B_{6}+2B_{7}+2B_{8}+B_{9} 
\right.\label{1vweights}
\\ 
 &&  \left. 
 +B_{10}-B_{11}-2B_{12}+B_{13}+B_{14}+B_{15}
 +B_{16}+B_{17}-B_{18}-B_{19}-B_{20}-B_{21} \right) 
 \nonumber
\ea
and for the superdiagrams with two vector lines 
\be
C_{\rm tot} = {g^4 \over 2} N^2(N^2-1)  \left( 2 C_{1} +2 C_{2} -2 C_{3} 
- 2 C_{4} + 
C_{5}- C_{6} +2 C_{7}  +2 C_{8} - C_{9}\right)  \: .
\label{2vweights}
\ee
 
Note that $C_{10}$ is zero due to Grassmann integration over the
$\theta$'s of the interaction vertices.
In general, the integration over the $\theta$ variables of the 
interaction vertices is greatly simplified by our choice of computing
the lowest component of the supercorrelation 
function~(\ref{4pointsuper}). This amounts to put to zero the $\theta$ variables
of the external insertions. 

The final result is a complicated combination of double, triple and 
quadruple integrals of convolutions of massless
scalar propagators and derivatives thereof.

The general strategy for computing the resulting integrals
is the following. Exploiting the 
conformal invariance of the full correlation function we send one of 
the external points
(say $x_1$) to infinity after  multiplying the correlator by
$(x_1^2)^{\Delta}$, where $\Delta$  is the conformal dimension of 
the operator inserted at $x_{1}$. Note that this simple prescription 
only works for operators (like ${\cal C}$) with 
protected conformal dimension ($\Delta=2$ in the case at hand), \ie 
independent of the coupling constant $g^{2}$.
For operators that acquire anomalous dimensions, 
one has to carefully subtract divergent
terms that behave like powers of $\log (x_1^2)$ as $x_1 \rightarrow \infty$.
In this limit the final answer will 
depend on the three variables $x_{23}^2$, $x_{24}^2$ and $x_{34}^2$. 
The complete $x_1$ dependence 
is dictated by conformal invariance and can be recovered after
the substitutions 
$x_{23}^2 \rightarrow x_{23}^2 x_{14}^2$, $x_{24}^2 \rightarrow 
x_{24}^2 x_{13}^2$ and 
$x_{34}^2 \rightarrow x_{34}^2 x_{12}^2$.

Taking the limit $x_1 \rightarrow \infty$ considerably 
simplifies the computation for two reasons. 
On the one hand it lowers the number of propagators to be integrated.
On the other hand, before taking the limit,
one can integrate by parts some of the derivatives
in the chiral propagators and make them act 
on propagators involving the point $x_1$. This increases 
the power of $x_1$ in the denominator making it larger than 4 with 
the consequence that terms generated in this way vanish
in the limit. This trick also
simplifies the $\theta$ integrals. As a result one is left with 
only double integrals except for the diagrams $B_{21}$ and $C_6$ 
where the remaining
integral is (naively) a triple one. A drawback of our 
procedure is  that the limit  $x_1 \rightarrow
\infty$ breaks many of the permutation symmetries of the integrand, 
thus diagrams that were related by rotations (or reflections) to one 
another must be separately computed. 
An important simplification arises by observing that 
in the limit $x_1 \rightarrow \infty$ certain  
linear combinations of diagrams contributing to~(\ref{1vweights}) 
and~(\ref{2vweights}) actually vanish. Precisely one finds 
$B_1=0$, $B_9=0$, $B_{15}-B_{18}=0$, $2 B_2-B_5-B_{20}=0$,
$B_3-2 B_6+B_{13}+B_{17}-B_{19}=0$, $C_1-C_3=0$, $C_2+C_8=0$, 
$2 C_4 - C_5 + C_9=0$. 
As already observed $C_{10}=0$ due to the $\theta$ integration even 
before taking the limit.

The whole result can be expressed in terms of only
two functions 
\be
g(i,j,k) = \int {dx_5 \over x_{i5}^2 x_{j5}^2 x_{k5}^2}
\label{defg}
\ee
and 
\be
f(i;j,k) = \int {dx_5 dx_6 \over x_{i5}^2 x_{j5}^2 x_{i6}^2 x_{k6}^2 
x_{56}^2} \; ,
\label{deff}
\ee
where in the notation used in~\cite{ehssw}
\be
f(i;j,k) \equiv f(i,j;i,k) \; .
\ee
A useful expression for the function $g$ is~\cite{bkrs}
\be
g(2,3,4) = {\pi^2 \over x_{23}^2} B \left({ x_{34}^2 \over  x_{23}^2},{ 
x_{24}^2
\over  x_{23}^2}\right) \; ,
\label{gfromB}
\ee
where $B(r,s)$ is a box-type integral given by
\be 
B(r,s) = \int \, d \beta_0 d \beta_1 d \beta_2 \: 
 \delta(1-\beta_0 - \beta_1 -\beta_2)  
\frac{1} 
{\beta_1 \beta_2 + r \beta_0 \beta_1  + s \beta_0 \beta_2 }
\; .
\label{Brs} 
\ee
The result of the integration in~(\ref{Brs}) can be expressed as a 
combination of logarithms and dilogarithms as follows
\ba
B(r,s) & = & 
{1 \over \sqrt{p}} \left \{ \ln (r)\ln (s)  - 
\left [\ln \left({r+s-1 -\sqrt {p} \over 2}\right)\right ]^{2} 
\right. \nonumber \\ 
&& \left. -2 {\rm Li}_2 \left({2 \over 1+r-s+\sqrt {p}}\right ) -
2 {\rm Li}_2 \left({2 \over 1-r+s+\sqrt {p}}\right )\right \} 
\; ,  \label{Brsf}
\ea
with ${\rm Li}_2 (z) = \sum_{n=1}^{\infty} {z^n\over n^2}$ 
and 
\be 
p = 1 + r^{2} + s^{2} - 2r - 2s - 2rs \; .  
\label{pdef}
\ee
The explicit expression of the function $f(i;j,k)$ has been obtained 
in~\cite{davyd} 
\be
f(i;j,k) = {\pi^{4}\over {x_{jk}^2}} 
\Phi^{(2)}\left({ x_{ij}^2 \over  x_{jk}^2},{x_{ik}^2 \over  x_{jk}^2}\right) 
\ee
where $\Phi^{(2)}(r,s)$ involves polylogarithms of up to fourth order. 
For the purpose of the present investigation, however, we only need
the following identities~\cite{ehssw}
\be
\Box_j f(i;j,k) =-4\pi^2 {1 \over x_{ij}^2 } g(i,j,k) \; ,
\label{boxf1}
\ee
\be
\Box_i f(i;j,k) =-4\pi^2 {x_{jk}^2 \over x_{ij}^2 x_{ik}^2} 
g(i,j,k) \; .
\label{boxf2}
\ee 
The latter identity has to be used in both directions to simplify the 
integrals appearing in the diagrams $A_2$, $B_7$, $B_8$ and $B_{11}$.
The triple integrals that appear in the computation of the 
diagrams $B_{21}$ and $C_6$ can be simplified
with the help of the relation
\be
\Box_3 \int {dx_5 \over x_{25}^5} \left[g(3,4,5)\right]^2  = 
-4\pi^2 {x_{24}^2 \over x_{34}^2 } \left[g(2,3,4)\right]^2
\label{tripi} \; .
\ee 
Eq.~(\ref{tripi}) can in turn be proved using the identity
\be
\Box_2 \left( x_{24}^2  \left[g(2,3,4)\right]^2 \right) = 
 x_{34}^2 \Box_3 \left[g(2,3,4)\right]^2 \; ,
\label{id2}
\ee
which can be verified by resorting to the explicit expression of 
$g(2,3,4)$ given by eq.~(\ref{gfromB}).

The total contributions of each of the three classes of diagrams of 
Figs.~\ref{diagrA}, \ref{diagrC} and~\ref{diagrB} is 
\be
A_{\rm tot} = {g^4 N^2(N^2-1)\over 8 (4\pi^{2})^{8}} \left\{{2 \over x_{34}^2} 
[f(3;2,4)+f(2;3,4)] + \left[g(2,3,4)\right]^2 \right\}\; ,
\label{Acontr}
\ee

\be
B_{\rm tot} = - {g^4 N^2(N^2-1) \over 4 (4\pi^{2})^{8}} {1 \over 4 
x_{34}^2}\left\{ \left[g(2,3,4)\right]^2
(x_{24}^2+x_{34}^2-x_{23}^2) -4f(3;2,4)-4f(4;2,3) \right\} \; ,
\label{Bcontr}
\ee

\be
C_{\rm tot} = {g^4 N^2(N^2-1) \over 2 (4\pi^{2})^{8}} {1 \over 4 
x_{34}^2} \left\{
x_{24}^2 \left[g(2,3,4)\right]^2 - 2 f(3;2,4) \right\}\; .
\label{Ccontr}
\ee

Summing the  above three contributions  yields 
\ba
&&  G(x_{2}, x_{3},x_{4}) \equiv
\lim_{x_{1}\rightarrow \infty } x_{1}^{4} G^{(H)}(x_{1},x_{2}, 
x_{3},x_{4}) \ = \ 
 {g^4 N^2 (N^2-1) \over 4 (4 \pi^2)^8 x_{34}^2} \ \times \nonumber \\
&&\left\{ f(2;3,4)+f(3;2,4)+f(4;2,3) \
+ \ {1 \over 4}
(x_{24}^2+x_{34}^2 + x_{23}^2) \left[g(2,3,4)\right]^2 \right\} \; .
\label{Ghtotal}
\ea
As already explained, the actual $x_1$ dependence is recovered by 
performing in the right hand side of~(\ref{Ghtotal}) the substitutions 
$x_{23}^2 \rightarrow x_{23}^2 x_{14}^2$, $x_{24}^2 \rightarrow 
x_{24}^2 x_{13}^2$ and $x_{34}^2 \rightarrow x_{34}^2 x_{12}^2$. 
The final result is in perfect agreement with the expression of the 
corresponding 
four-point function of hypermultiplet bilinears computed by means 
of ${\cal N}$=2 harmonic superspace techniques in~\cite{ess}.

\section{Logarithms and anomalous dimensions}

As in previous computations at order 
$g^{2}$~\cite{ehssw,bkrs} and at the one-instanton
level~\cite{bkrs}, the function~(\ref{Ghtotal}) shows the 
expected short-distance logarithmic 
behaviour. Indeed at short distances (\ie in the limit in which
pairs of points are taken to coincide) one finds linear and 
quadratic logarithms that are not incompatible with the finiteness of 
${\cal N}=4$ SYM in the superconformal phase. 

To illustrate this point in a simple fashion we will follow 
closely~\cite{bkrs} 
and consider as an example the 
two-point function of a primary operator of scale dimension $\Delta$ 
\begin{equation}
\langle {\cal O}^\dagger_\Delta (x) {\cal O}_\Delta (y) \rangle   = 
{A_\Delta 
\over 
(x-y)^{2\Delta}} \; ,
\label{exact}
\end{equation}
where $A_\Delta$ is an overall normalisation constant possibly 
depending on the subtraction scale $\mu$. Now suppose that 
$\Delta = \Delta^{^{(0)}} + \gamma$, \ie the operator under 
consideration 
has an anomalous dimension.  In  perturbation theory $\gamma=\gamma 
(g^{2})$ 
is expected to be small and to admit an expansion  in the
coupling constant $g^{2}$ 
\be
\gamma (g) = \gamma_{1} g^{2} + \gamma_{2} g^{4} + \ldots \; .
\label{gammaexp}
\ee
The perturbative expansion of~(\ref{exact}) in powers of $g^{2}$ yields  
\ba
&&\langle {\cal O}^\dagger_\Delta (x) {\cal O}_\Delta (y)\rangle
= { a_\Delta \over (x-y)^{2\Delta^{^{(0)}}}}  \ \times \label{defdelta}
\\ 
&&\left( 1 - g^{2} \gamma_{1}  \log 
[\mu^2 (x-y)^2] + 
g^{4} \left\{ {1 \over 2} \gamma_{1}^2 ( \log [\mu^2 (x-y)^2] )^2
- \gamma_{2} \log [\mu^2 (x-y)^2] \right\} 
 + \ldots \right) \;  \nonumber ,
\ea
where after renormalisation 
we have set $A_\Delta = a_\Delta \mu^{-2\gamma}$.
Thus, although the exact expression~(\ref{exact}) is conformally 
invariant, 
at each  order in $g^{2}$ eq.~(\ref{defdelta}) contains powers of
logarithms that seem to even violate  
scale invariance. Similar considerations apply to generic  
$n$-point Green functions as well. 
 
Assuming the validity of the OPE, a four-point 
function can be expanded in the $s$-channel in the form
\begin{equation}
\langle {\cal Q}_{A}(x) {\cal Q}_{B}(y) {\cal Q}_{C}(z) {\cal 
Q}_{D}(w) 
\rangle  = 
\sum_K 
{C_{CD}{}^K (z-w,\partial_{w})\over (z-w)^{\Delta_C + 
\Delta_D-\Delta_K}} 
\langle {\cal Q}_{A}(x) {\cal Q}_{B}(y) {\cal O}_{K}(w) \rangle \; ,
\label{doubleope}
\end{equation}
where $K$ runs over a  
complete set of primary operators. Descendants 
are implicitly taken into account by the presence of derivatives in 
the 
Wilson coefficients, $C$'s. An expansion like~(\ref{doubleope}) is 
valid 
in the other two channels  as well. To simplify formulae we assume 
that 
${\cal Q}_{A},{\cal Q}_{B},{\cal Q}_{C},{\cal Q}_{D}$ are protected 
operators,~\ie they have no 
anomalous dimensions. In general the operators ${\cal O}_{K}$ may 
have anomalous dimensions, 
$\gamma_K$, so that $\Delta_K=\Delta^{^{(0)}}_K + \gamma_K$, where 
$\Delta^{^{(0)}}_K$ is the tree-level scale dimension. Similarly 
$C_{IJ}{}^K = C^{^{(0)}}_{IJ}{}^K + \eta_{IJ}{}^K$, with 
$\eta_{IJ}{}^K$ the perturbative correction to the OPE coefficients.

The three-point function in the right hand side of eq.~(\ref{doubleope})
is determined by conformal invariance to be of the form
\begin{equation}
\langle {\cal Q}_{A}(x) {\cal Q}_{B}(y){\cal O}_{K}(z)
\rangle  = 
{ C_{ABK}(g^{2})
\over (x-y)^{\Delta_A + \Delta_B-\Delta_K}
(x-z)^{\Delta_A - \Delta_B+\Delta_K}
(y-z)^{\Delta_B - \Delta_A+\Delta_K} } \; .
\label{3pointr}
\end{equation}
Eq.~(\ref{3pointr}), like the two-point function~(\ref{defdelta}), 
can be expanded in 
power series in $g^{2}$ giving rise to logarithmic terms.
From these formulae one can extract the corrections
to both the OPE coefficients and the anomalous dimensions
of the operators ${\cal O}_{K}$. 
The same procedure applies  also to  derivatives of the three- and 
four-point functions as well as to their limits for $x_{1} \rightarrow 
\infty$. 

Let us analyze the function $G^{(H)}$ in 
the limit $x_{4}\rightarrow x_{3}$ that exposes 
the $s$-channel. Only operators in the singlet of 
the manifest $SU(3)\times U(1)$ may be exchanged \cite{bkrs}. 
Barring the identity operator, that is 
clearly not renormalized, the leading contribution comes from the operator
\begin{equation} 
{\cal K}_{{\bf 1}} = \frac{1}{3} : \tr  (\varphi^{i} \varphi^{i}) : 
\; , 
\label{k1}
\end{equation}
that has naive dimension $\Delta^{^{(0)}} = 2$ and is lowest component
of the long Konishi supermultiplet. 

In the case under consideration, the relative complexity of the 
explicit expression for $f(i;j,k)$
makes it difficult to directly analyse the short distance behaviour of 
the function~(\ref{Ghtotal}). Thus we find it more convenient to
compute $\Box_2 G(x_{2}, x_{3},x_{4})$, 
which can be expressed in terms of only the much simpler 
function $g(i,j,k)$.
Using eqs.~(\ref{boxf1}), (\ref{boxf2}) and~(\ref{id2}), one obtains
\ba
\Box_2 G(x_{2}, x_{3},x_{4}) &=& 
{g^4 N^2 (N^2-1) \over 8 (4 \pi^2)^8 } \left[
\sum_{j=2}^4 \partial_j g(2,3,4) \cdot \partial_j g(2,3,4)
\right.  \nonumber \\
&-&3 (4 \pi^2)\left. \left({1 \over x_{23}^2 x_{34}^2}+
{1 \over x_{24}^2 x_{34}^2}+
{1 \over x_{23}^2 x_{24}^2}\right) g(2,3,4) \right] \; .
\label{BoxGh}
\ea 

The leading behaviour 
of~(\ref{BoxGh}) in the limit $x_{4}\rightarrow x_{3}$ is given by  
\be
\Box_2 G(x_2,x_3,x_4) ~~ \longrightarrow 
\raisebox{-12pt}{\hspace*{-28pt} $\begin{scriptstyle}x_4 \to x_3
\end{scriptstyle}$}~
{g^4 N^2 (N^2-1) \over 16 (4 \pi^2)^6 } \ {1 \over x_{23}^4 x_{34}^2} 
\left[ 3 \log \left({x_{34}^2 \over x_{23}^2} \right) - 5\right] \; .
\label{sing}
\ee

We now compare eq.~(\ref{sing}) with the result of the OPE analysis of the 
order $g^{4}$ contribution to the function $\Box_2 G(x_{2}, x_{3},x_{4})$.  
The latter yields
\be
\Box_2 G(x_{2}, x_{3},x_{4}) ~~ \longrightarrow 
\raisebox{-12pt}{\hspace*{-28pt} $\begin{scriptstyle}x_4 \to x_3
\end{scriptstyle}$}~  
{g^4 (N^2-1) \over (4 \pi^2)^6  x_{23}^4 x_{34}^2} 
\left[ a_{0} \gamma_{1}^{2}  \log \left({x_{34}^2 \over x_{23}^2} 
\right)
+a_{0}(\gamma_{1}^{2}+ 2 \gamma_{2}) +2 a_{1}\gamma_{1} \right] 
\; ,
\label{opeboxg0}
\ee
where the coefficients $a_{0}$ and $a_{1}$ are given by
\be 
a_{0} = {1 \over 3} (4 \pi^{2})^{2} \ , \qquad \qquad 
a_{1} = -N \pi^{2} \; .
\ee
They represent the tree-level contribution of the Konishi scalar ${\cal 
K}_{\bf 1}$ 
to the $s$-channel expansion of the function
$G^{(H)}$ and its finite one-loop correction.
The coefficients $a_{0}$ and $a_{1}$
are determined by the one-loop analysis performed in \cite{bkrs} 
that gives
\ba
G(x_2,x_3,x_4)\vert_{1-{\rm loop}} &\longrightarrow 
\raisebox{-12pt}{\hspace*{-28pt} $\begin{scriptstyle}x_4 \to x_3
\end{scriptstyle}$}&
{g^2 (N^2-1) \over (4 \pi^2)^6 } \ {1 \over x_{23}^2 x_{34}^2} 
\left[ a_{0} {\gamma_{1}\over 2} 
\log \left({x_{34}^2 \over x_{23}^2} \right) + a_{1} \right] \nonumber \\
&& = {g^2 (N^2-1) \over (4 \pi^2)^6 } \ {1 \over x_{23}^2 x_{34}^2} \pi^{2} N
 \left[ {1\over 2} 
\log \left({x_{34}^2 \over x_{23}^2} \right) - 1 \right] \; .
\label{sing1loop}
\ea
We stress that the quadratic logarithmic term in  $G^{(H)}$ contributes a 
linear logarithmic term to $\Box_2 G^{(H)}$ in eq.~(\ref{opeboxg0}.

Comparison of eq.~(\ref{sing}) with eq.~(\ref{opeboxg0}) shows that
the coefficient of the logarithmic term is related to the square of the 
one-loop 
anomalous dimension of ${\cal K}_{\bf 1}$, and agrees with the known value 
$\gamma_1 = {3 N \over 16 \pi^2 }$ \cite{ans}. 
The the constant term corresponds to a  combination of the one- and 
two-loop
anomalous dimensions and the known one-loop correction to the OPE
coefficient. For the $g^{4}$ contribution to the anomalous 
dimension of the Konishi supermultiplet one obtains 
\be
\gamma_{\rm 2-loop} \  = \ \gamma_2 \;  g^{4} \ = \ 
- {3 \ g^{4} N^2 \over 16 \  (4 \pi^2 )^2} \; .
\label{gamma2}
\ee

The analysis of the other two non-singlet channels ($x_{24} 
\rightarrow 0$ and $x_{23} \rightarrow 0$) is more subtle.
The lowest dimensional operators that  can be exchanged in both 
channels belong to the ${\bf 105}$ and to the ${\bf 84}$
representations of the $SU(4)$ R-symmetry. Single- and double-trace
operators of dimension four in the ${\bf 105}$ are expected to be 
protected~\cite{af,skiba}. In order to disentangle the two $SU(4)$ 
representations it is necessary to consider another four-point 
function. We find it convenient to analyse the four-point 
correlator $G^{(V)}$, defined in~(\ref{all4point}).
One-loop~\cite{ehssw}, one-instanton~\cite{bgkr,bkrs}
and two-loop~\cite{ess} computations give $s G^{(V)} = G^{(H)}$. 
The OPE analysis of the $g^{4}$ contribution to $G^{(V)}$,
as it is obtained from the above functional relation, confirms the 
non-renormalisation of the  ${\bf 105}$ operators.

The two possible operators of 
naive dimension four in the ${\bf 84}$ 
mix at one-loop~\cite{bkrs}. One of them, 
${\cal K}_{{\bf 84}}$ is a 
superconformal descendant of of ${\cal K}_{{\bf 1}}$
and as such has the same anomalous dimension.
A careful OPE analysis,
combined with the symmetry of the factor in brackets in the right 
hand side of eq.~(\ref{Ghtotal}) under the exchange of $x_{2}$ and $x_{4}$, 
implies that this operator saturates the logarithmic 
behaviour in this channel. We thus conclude that the operator 
$\hat{\cal D}_{{\bf 84}}$, identified in~\cite{bkrs} as the 
combination orthogonal to ${\cal K}_{{\bf 84}}$, is 
protected also at order $g^{4}$. This is in agreement with the fact that 
it belongs to a supermultiplet satisfying a generalized shortening 
conditions~\cite{dp,fz,afsz,fsok,fertmr,gunaydin}.

\section{Comments}

Let us briefly comment on the bearing of the results of this paper.
First of all the very fact that it was possible to compute in 
closed form
a four-point function of protected composite operators at order $g^4$
shows that ${\cal N}$=4 SYM is both non-trivial and calculable.
No off-shell approach is known that preserves ${\cal N}$=4 
supersymmetry. At the quantum level, one has to resort either to
the ${\cal N}$=1 superspace approach pursued here or to the less 
familiar
${\cal N}$=2 harmonic superspace approach pursued in~\cite{ess}. 
The coincidence of all known results in the two approaches
gives independent support 
to some ${\cal N}$=2 harmonic superspace identities~\cite{nilpot}, 
based on the bonus symmetry proposed 
in~\cite{intri}, that led to drastic simplifications in the 
computations reported in~\cite{ess}.

The introduction of
supersymmetric mass-terms gives rise to interesting, in some cases 
confining,
theories that
can be handled with ${\cal N}$=1 superfield techniques. 
Alternatively one might
consider other finite ${\cal N}$=1 gauge theories some of which are 
conjectured
to be dual to type IIB superstring on $AdS_5\times S^5/\Gamma$, where
$\Gamma$ is a discrete subgroup of an $SU(3)$ subgroup of the $SU(4)$
R-symmetry. These gauge theories govern the dynamics of the light 
degrees of freedom of stacks of coincident
D3-branes at special orbifold singularities~\cite{vafa}.
Other ${\cal N}$=1 finite theories can be obtained by perturbing 
${\cal N}$=2
theories by mass-terms. This is the case of D3-branes at generalized
conifold singularities~\cite{kv}, whose supergravity dual replaces 
$S^5$ with less trivial Einstein spaces~\cite{andria}. 
In all such cases the ${\cal N}$=1 formalism pursued in this
paper might allow one to compute correlation functions of protected 
composite
operators. By OPE one would then
extract anomalous dimensions and couplings of unprotected
operators such as those belonging to Konishi-like long 
supermultiplets.
These are expected to be dual to genuine string excitations and as 
such should
decouple from the operator algebra in the supergravity limit, 
that is dual to the large $N$ and strong 't Hooft coupling limit.
The fact that the one-loop anomalous dimension of the Konishi 
multiplet
has been known for some time~\cite{ans}
to be positive was somewhat reassuring in this respect.
The OPE analysis of the two-loop computations
confirm the one-loop result, 
but at the same time yields a negative two-loop contribution
to the anomalous dimension. This result certainly
requires further investigation and some cross-checks~\cite{prep}.
Another related issue is the r\^ole of multi-trace operators that are 
expected to
be dual to multiparticle states in the AdS description. We have shown  
and
confirmed in the present paper that their mixing with single trace
operators is not suppressed in the large $N$ limit at finite 't Hooft
coupling. In fact protected operators belonging to supermultiplets 
that satisfy 
generalized shortening conditions~\cite{dp,fz,afsz,fsok,fertmr,gunaydin}
are typically mixtures of single- and multi-trace operators. One 
would like to
clarify their r\^ole in the operator algebra in relation to the ``string 
exclusion
principle" that is expected to drastically reduce the spectrum of AdS
excitations at finite ``radius", \ie at finite $N$~\cite{maldstrom}.

\vspace*{1cm}

{\large{\bf Acknowledgements}}
\vspace*{0.3cm}

\noindent
We would like to thank E.~D'Hoker,   
S.~Ferrara, M.~Gunaydin and J.F.~Morales for useful discussions. 
Ya.S.S. and S.K. would like to thank 
the Physics Department and I.N.F.N. Section at Universit\`a di Roma 
``Tor Vergata'' for hospitality and financial support.


\begin{thebibliography}{123}

        
\bibitem{jm}{J. Maldacena, ``The large $N$ limit of superconformal
field theories and supergravity'', {\it Adv. Theor. Math. Phys.} {\bf 2} 
(1998) 231, {\tt  hep-th/9711200}.}

\bibitem{gkp}{S.S. Gubser, I.R. Klebanov and A.M. Polyakov, ``Gauge
theory correlators from non-critical string theory'', {\it
Phys. Lett.} {\bf B428} (1998) 105.}

\bibitem{wads}{E. Witten, ``Anti De Sitter space and holography'',
{\it Adv. Theor. Math. Phys.} {\bf 2} (1998) 253.}

\bibitem{agmoo}{O. Aharony, S.S. Gubser, J. Maldacena, H. Ooguri and Y. 
Oz, ``Large $N$ field theories, string theory and gravity'', 
{\it Phys. Rept.} {\bf 323} (2000) 183, {\tt hep-th/9905111}.}

\bibitem{lmrs}{S. Lee, S. Minwalla, M. Rangamani and N. Seiberg,
``Three-point functions of chiral operators in D=4 ${\cal N}$=4 SYM
at large $N$'',  
{\it Adv. Theor. Math. Phys.} {\bf 2} (1998) 697, {\tt 
hep-th/9806074}.}

\bibitem{dfs}{E. D'Hoker, D.Z. Freedman and W. Skiba, ``Field theory
tests for correlators in the AdS/CFT correspondence'', {\it Phys. 
Rev.} {\bf D59} (1999) 045008, {\tt hep-th/9807098}.}

\bibitem{skiba}{W. Skiba, ``Correlators of short multi-trace operators
in ${\cal N}$=4 supersymmetric Yang--Mills theory'', {\it Phys. Rev.}
{\bf D60} (1999) 105038, {\tt hep-th/9907088}.}

\bibitem{extremal}{E. D'Hoker, D.Z. Freedman, S.D. Mathur, A. Matusis 
and L. Rastelli, ``Extremal correlators in the AdS/CFT
correspondence'', {\tt hep-th/9908160}.}

\bibitem{bkextr}{M. Bianchi and S. Kovacs, ``Non-renormalisation of
extremal correlators in ${\cal N}$=4 SYM theory'', {\it Phys. Lett.}
{\bf B468} (1999) 102, {\tt hep-th/9910016}.}

\bibitem{ehsswextr}{B. Eden, P.S. Howe, C. Schubert, E. Sokatchev and 
P.C. West,  ``Extremal correlators in four-dimensional SCFT'',
{\it Phys. Lett.} {\bf B472} (2000) 323, 
{\tt hep-th/9910150}.}

\bibitem{epv}{J. Erdmenger and M. Perez-Victoria,
``Non-renormalisation of next-to-extremal correlators in ${\cal N}$=4
SYM and the AdS/CFT correspondence'', {\tt hep-th/9912250}.}

\bibitem{ehssw}{B. Eden, P.S. Howe, C. Schubert, E. Sokatchev and 
P.C. West, ``Four-point functions in ${\cal N}$=4 supersymmetric
Yang--Mills theory at two loops'', {\it Nucl. Phys.} {\bf B557} (1999)
355, {\tt hep-th/9811172}; 
``Simplifications of four-point functions in ${\cal N}$=4
supersymmetric Yang--Mills theory at two loops'', {\it Phys. Lett.} 
{\bf B466} (1999) 20, {\tt hep-th/9906051}.}  

\bibitem{bkrs}{M. Bianchi, S. Kovacs, G.C. Rossi and Ya.S. Stanev, 
``On the logarithmic behaviour in ${\cal N}$=4 SYM theory'', 
{\it J. High Energy Phys.} {\bf 08} (1999) 020, {\tt hep-th/9906188}.}

\bibitem{gg} {M.B. Green and M. Gutperle, ``Effects of 
D-instantons'', 
{\it Nucl. Phys.} {\bf B498} (1997) 195, {\tt hep-th/9701093};
``D-particle bound states and the D-instanton measure'', 
{\it J. High Energy Phys.} 
{\bf 9801} (1998) 005, {\tt hep-th/9711107}; \\
M.B. Green and P. Vanhove, ``D-instantons, strings and M-theory'', 
{\it Phys. Lett.} {\bf B408} (1997) 122, {\tt hep-th/9704145}; \\
M.B. Green,  M. Gutperle and H. Kwon, ``Sixteen-fermion and related
terms in M-theory on $T^2$'',  
{\it Phys. Lett.} {\bf B421} (1998) 149, {\tt hep-th/9710151} .}

\bibitem{bg}{T. Banks and M.B. Green, ``Non-perturbative effects in
$AdS_5\times S^5$ string theory and d=4 SUSY Yang--Mills'', 
{\it J. High Energy Phys.} {\bf 05} (1998) 002, {\tt hep-th/9804170}.}

\bibitem{bgkr}{M. Bianchi, M.B. Green, S. Kovacs and G.C. Rossi, 
``Instantons in supersymmetric Yang--Mills and D-instantons in IIB
superstring theory'', 
{\it J. High Energy Phys.} {\bf 08} (1998) 013, {\tt hep-th/9807033}.}

\bibitem{como}{M. Bianchi and S. Kovacs, ``Yang--Mills instantons
vs. type IIB D-instantons'', {\tt hep-th/9811060}.}

\bibitem{dhkmv}{N. Dorey, V.V. Khoze, M.P. Mattis and S. Vandoren,
``Yang--Mills instantons in the large-$N$ limit and the AdS/CFT 
correspondence'', 
{\it Phys. Lett.} {\bf B442} (1998) 145, {\tt hep-th/9808157}; \\
N. Dorey, T.J. Hollowood, V.V. Khoze, M.P. Mattis and S. Vandoren,
``Multi-instantons and Maldacena's conjecture'',  
{\it J. High Energy Phys.} {\bf 06} (1999) 023, {\tt hep-th/9810243}; 
``Multi-instanton calculus and the AdS/CFT correspondence in ${\cal
N}$=4 superconformal field theory'', {\it Nucl. Phys.} {\bf B552}
(1999) 88, {\tt hep-th/9901128}.}

\bibitem{dmmrope}{E. D'Hoker, D.Z. Freedman, S.D. Mathur, A. Matusis
and L. Rastelli, ``Operator product expansion of ${\cal N}$=4 SYM and
the 4-point functions of supergravity'', {\tt hep-th/9911222}.}

\bibitem{arfr2}{G. Arutyunov and S. Frolov, ``Four-point functions of
lowest weight CPOs in ${\cal N}$=4 SYM$_4$ in supergravity
approximation'', {\tt hep-th/0002170}.}

\bibitem{arfr1}{G. Arutyunov and S. Frolov, ``Scalar quartic couplings in
type IIB supergravity on $AdS_5\times S^5$'', {\tt hep-th/9912210}.}

\bibitem{ehpsw}{B. Eden, P.S. Howe, A. Pickering, E. Sokatchev and
P.C. West, ``Four-point functions in ${\cal N}$=2 superconformal 
theories'', {\tt hep-th/0001138}.}

\bibitem{ess}{B. Eden, C. Schubert and E. Sokatchev, ``Three-loop
four-point correlator in ${\cal N}$=4 SYM'', {\tt hep-th/0003096}.}

\bibitem{n4sym}{L. Brink, J. Scherk and J.H. Schwarz, ``Supersymmetric
Yang--Mills theories'', 
{\it Nucl. Phys.} {\bf B121} (1977) 77; \\
F. Gliozzi, D.I. Olive and J. Scherk, ``Supersymmetry , supergravity
and the dual spinor model'', {\it Nucl. Phys.} {\bf B122} 
(1977) 253.} 

\bibitem{finite}{S.~Ferrara and B.~Zumino, ``Supergauge invariant
Yang--Mills theories'', 
{\it Nucl. Phys.} {\bf B79} (1974) 413; \\
M.~Grisaru, M.~Ro\v{c}ek and W.~Siegel, ``Zero value of the three-loop
$\beta$ function in ${\cal N}$=4 supersymmetric Yang--Mills theory'', 
{\it Phys. Rev. Lett.} {\bf 45} (1980) 1063; \\ 
W.E.~Caswell and D.~Zanon, ``Zero three-loop
beta function in the ${\cal N}$=4 supersymmetric Yang--Mills 
theory'', {\it Nucl. Phys.} {\bf B182} (1981) 125.}

\bibitem{kov}{S. Kovacs, {``A perturbative re-analysis of ${\cal N}$=4 
supersymmetric Yang--Mills theory''}, 
{\tt hep-th/9902047}; {\it ${\cal N}$=4 
supersymmetric Yang--Mills theory and the 
AdS/SCFT correspondence}, PhD thesis, {\tt hep-th/9908171}.}

\bibitem{psz}{S. Penati, A. Santambrogio and D. Zanon, ``Two-point
functions of chiral operators in ${\cal N}$=4 SYM'', {\it J. High
Energy Phys.} {\bf 12} (1999) 006, {\tt hep-th/9910197}.}

\bibitem{davyd}{N.I. Ussyukina and A.I. Davydychev, ``Exact results
for three and four point ladder diagrams with an arbitrary number of 
rings'', {\it Phys. Lett.} {\bf B298} (1993) 363.}

\bibitem{ans}{D. Anselmi, D.Z. Freedman, M.T. Grisaru and A.A. 
Johansen, ``Universality of the operator product expansion of
SCFT$_4$'', 
{\it Phys. Lett.} {\bf B394} (1997) 329, {\tt hep-th/9608125};
``Nonperturbative formulas for central functions of supersymmetric
gauge theories'',  {\it Nucl. Phys.} {\bf B526} (1998) 543, 
{\tt hep-th/9708042}; \\
D. Anselmi, J. Erlich, D.Z. Freedman and A. Johansen, ``Positivity
constraints on anomalies in supersymmetric gauge theories'', 
 {\it Phys. Rev.} {\bf D57} (1998) 7570, {\tt hep-th/9711035}; \\
D. Anselmi, ``The ${\cal N}$=4 quantum conformal algebra'', 
{\it Nucl. Phys.} {\bf B541} (1999) 369, {\tt hep-th/9809192}.}

\bibitem{af}{L. Andrianopoli and S. Ferrara, ``KK excitations on
$AdS_5\times S^5$ as ${\cal N}$=4 1`primary' superfields'', {\it Phys. 
Lett.} {\bf B430} (1998) 248, {\tt hep-th/9803171}; 
``On short and long $SU(2,2|4)$ multiplets in
the AdS/CFT correspondence'', {\it Lett. Math. Phys.} {\bf 48} (1999)
145, {\tt hep-th/9812067}.}

\bibitem{dp}{V.K. Dobrev and V.B. Petkova, ``All positive energy
unitary irreducible representations of extended conformal
supersymmetry'', {\it Phys. Lett.} {\bf 162B} (1985) 127.}

\bibitem{fz}{S. Ferrara, talk given at ``Strings '99''; \\ 
S. Ferrara and A. Zaffaroni, ``Superconformal field theories,
multiplet shortening and the $AdS_5 \times S^5$ correspondence'', 
{\tt hep-th/9908163}.}

\bibitem{afsz}{L. Andrianopoli, S. Ferrara, E. Sokatchev and
B. Zupnik, ``Shortening of primary operators in ${\cal N}$-extended
SCFT$_4$ and harmonic-superspace analyticity'', {\tt hep-th/9912007}.}

\bibitem{fsok}{S. Ferrara and E. Sokatchev, ``Short representations of
SU(2,2$|{\cal N}$) and harmonic superspace analyticity'', 
{\tt hep-th/9912168}.}

\bibitem{fertmr}{S. Ferrara, ``Superspace representations of
SU(2,2,$|{\cal N}$) superalgebras and multiplet shortening'', 
{\tt hep-th/0002141}.}

\bibitem{gunaydin}{M. Gunaydin, D. Minic and M. Zagerman, ``Novel
supermultiplets of $SU(2,2|4)$ and the $AdS_5$/CFT$_4$ duality'', 
{\it Nucl. Phys.} {\bf B544} (1999) 737, {\tt hep-th/9810226}; ``4d
doubleton conformal theories, CPT and IIB string on $AdS_5\times
S^5$'', {\it Nucl. Phys.} {\bf B534} (1998) 96, {\tt hep-th/9806042}; 
Erratum-ibid. {\bf B538} (1999) 531; \\
P. Claus, M. Gunaydin, R. Kallosh, J. Rahmfeld and  Y. Zunger,
``Supertwistors as quarks of $SU(2,2|4)$'', {\it J. High Energy Phys.}
{\bf 05} (1999) 019, {\tt hep-th/9905112}.}

\bibitem{nilpot}{P.S. Howe, C. Schubert, E. Sokatchev and P.C. West,
`` Explicit construction of nilpotent covariants in ${\cal N}$=4
SYM'', {\tt hep-th/9910011}.}

\bibitem{intri}{K. Intriligator, ``Bonus symmetry of ${\cal N}$=4
super Yang--Mills correlation functions via AdS Duality'', 
{\it Nucl. Phys.} {\bf B551} (1999) 575, {\tt hep-th/9811047}; \\
K. Intriligator and W. Skiba, ``Bonus symmetry and the operator
product expansion of ${\cal N}$=4 super Yang--Mills'', 
{\it Nucl. Phys.} {\bf B559} (1999) 165, 
{\tt hep-th/9905020}.}

\bibitem{vafa}{S. Kachru and E. Silverstein, 
``4-D conformal Theories  and Strings on Orbifolds'', 
{\it Phys. Rev. Lett.} {\bf 80} (1998) 4855, hep-th/9802183; \\ 
M. Bershadsky, Z. Kakushadze and C. Vafa, ``String  
expansion as large N expansion of gauge theories'', {\it Nucl. Phys.}
{\bf B523} (1998) 59, hep-th/9803076; \\ 
Z. Kakushadze, ``Gauge theories from orientifolds and large N 
limit'', {\it Nucl. Phys.} {\bf B529} (1998) 157, hep-th/9803214; \\
A.Lawrence, N.Nekrasov and C.Vafa, ``On Conformal Field Theories in 
Four Dimensions'', {\it Nucl. Phys.} {\bf B533} (1998) 199, hep-th/9803015; \\
M. Bershadsky and A. Johansen, ``Large $N$ Limit of Orbifold Field  
Theories'', {\it Nucl. Phys.} {\bf B536} (1998) 141, hep-th/9803249.}

\bibitem{kv}{I. Klebanov and E. Witten, ``Superconformal field theory
on three-branes at a Calabi--Yau singularity'', {\it Nucl. Phys.} {\bf
B536} (1998) 199, {\tt hep-th/9807080}.}

\bibitem{andria}{A. Ceresole, G. Dall'Agata, R. D'auria and
S. Ferrara, ``Spectrum of type IIB supergravity on $AdS_5 \times
T^{11}$: predictions on ${\cal N}$=1 SCFT's'', {\it Phys. Rev.} {\bf
D61} (2000) 066001, {\tt hep-th/9905226}; ``Superconformal field
theories from IIB spectroscopy on $AdS_5\times T^{11}$'', {\it
Class. Quant. Grav.} {\bf 17} (2000) 1017, {\tt hep-th/9910066}.}

\bibitem{prep}{M. Bianchi, S. Kovacs, G.C. Rossi and Ya.S. Stanev, in
preparation.}

\bibitem{maldstrom}{J.~Maldacena and A.~Strominger, ``$AdS_3$ Black
Holes and a Stringy Exclusion Principle'', {\it J. High Energy Phys.}
{\bf 12} (1998) 005, {\tt hep-th/9804085}; \\ 
A.~Giveon, D.~Kutasov and N.~Seiberg, ``Comments on String Theory
on $AdS_3$'', {\it Adv. Theor. Math. Phys.} {\bf 2} 
(1998) 733, {\tt hep-th/9806194}.}



\end{thebibliography}
\end{document}